\documentstyle[12pt]{article}
\newcommand{\Ebf}{\mbox{\boldmath $E$}}

\newcommand{\Hbf}{\mbox{\boldmath $H$}}

\newcommand{\zbf}{\mbox{\boldmath $z$}}

\newcommand{\nabf}{\mbox{\boldmath $\nabla$}}
\newcommand{\rhobf}{\mbox{\boldmath $\rho$}}
\newcommand{\TM}{{\rm TM}}

\newcommand{\ph}{{\rm ph}}

\newcommand{\pa}{\partial}

\newcommand{\X}{{\rm X}}
\newcommand{\pisr}{{\sqrt{\pi}}}

\newcommand{\text}{\rm}

\newcommand{\drm}{{\rm d}}
\newcommand{\grm}{{\rm g}}

\newcommand{\ug}{ \; = \; }
\newcommand{\ugg}{ \ = \ }

\newcommand{\infi}{\infty}

\newcommand{\la}{\lambda}

\newcommand{\bb}{\begin{equation}}
\newcommand{\ee}{\end{equation}}
\newcommand{\bega}{\begin{eqnarray}}
\newcommand{\ega}{\end{eqnarray}}
\newcommand{\begae}{\begin{eqnarray*}}
\newcommand{\egae}{\end{eqnarray*}}

\newcommand{\h}{\hspace*{4ex}}
\newcommand{\dis}{\displaystyle}

\newcommand{\be}{\beta}

\newcommand{\th}{\theta}

\newcommand{\om}{\omega}

\newcommand{\cent}{\centerline}
\newcommand{\vs}{\vspace*}

\begin{document}
 
\baselineskip 0.65cm
 
\begin{center}
 
{\large {\bf Superluminal Localized Solutions to Maxwell Equations
propagating along a waveguide: The finite-energy case}$^{\: (\dag)}$}
\footnotetext{$^{\: (\dag)}$ 
Work partially supported by CAPES (Brazil), and by MIUR/MURST and INFN
(Italy). \ E-mail address for contacts: recami@mi.infn.it}
 
\end{center}
 
\vs{5mm}
 
\cent{ Michel Zamboni Rached, }
 
\vs{0.2 cm}
 
\centerline{{\em D.M.O., Faculty of Electrical Engineering, UNICAMP, Campinas,
SP, Brasil.}}
 
\vs{0.7 cm}

\cent{ Flavio Fontana }
 
\vs{0.2 cm}
 
\cent{{\em R\&D Sector, Pirelli Labs, Milan, Italy}}
 
\vs{0.2 cm}

\centerline{\rm and}
 
\vs{0.3 cm}

\cent{ Erasmo Recami }
 
\vs{0.2 cm}
 
\cent{{\em Facolt\`a di Ingegneria, Universit\`a statale di Bergamo,
Dalmine
(BG), Italy;}}
\cent{{\em INFN---Sezione di Milano, Milan, Italy; \ {\rm and}}}
\cent{{\em CCS, State University of Campinas,
Campinas, S.P., Brasil.}}

\vs{1. cm}
 
\
 
{\bf Abstract  \ --} \ In a previous paper we have shown localized
(non-evanescent) solutions to Maxwell equations to exist, which propagate
without distortion with Superluminal speed along normal-sized waveguides,
and consist in trains of ``X-shaped" beams.  Those solutions possessed
therefore infinite energy. In this note we show how to obtain, by
contrast,
{\em finite}-energy solutions, with the same localization and
Superluminality properties.
 
\
 
{\em PACS nos.}: \ 41.20.Jb ; \ 03.50.De ; \ 03.30.+p ; \ 84.40.Az ; \ 
42.82.Et \ .

{\em Keywords\/}: Wave-guides; Localized solutions to Maxwell equations;
Superluminal waves; Bessel beams; Limited-dispersion beams; Finite-energy
waves;
Electromagnetic wavelets; X-shaped waves; Evanescent waves;
Electromagnetism;
Microwaves; Optics; Special relativity; Localized acoustic waves;
Seismic waves; Mechanical waves; Elastic waves; Guided gravitational
waves.

\newpage

{\bf 1. -- Introduction: Localized solutions to the wave equation}\\
 
\h Already in 1915 Bateman[1] showed that Maxwell equations admit
(besides
of
the ordinary solutions, endowed in vacuum with speed $c$) of
wavelet-type solutions, endowed in vacuum with group-velocities $0 \le v
\le c$. But Bateman's work went practically unnoticed. Only few authors,
as Barut et al.[2], followed such a research line; incidentally, Barut et
al. constructed even a wavelet-type solution travelling with Superluminal
group-velocity[3] $v > c$.
 
\h In more recent times, however, many authors discussed the fact that
all
(homogeneous) wave equations admit solutions with $0 < v < \infty$: \
see,
e.g., refs.[4]. Most of  those authors confined themselves to investigate
(sub- or Super-luminal) {\em localized} non-dispersive solutions in
vacuum:
namely, those solutions that were called ``undistorted progressive waves"
by Courant \& Hilbert. Among localized solutions, the most interesting
appeared to be the so-called ``X-shaped" waves, which ---predicted to
exist
even by Special Relativity in its extended version[5]--- had been
mathematically constructed by Lu \& Greenleaf[6] for acoustic waves, and
by
Ziolkowski et al.[6], and later Recami[6], for electromagnetism. \ Let us
recall that such ``X-shaped" localized solutions are Superluminal
(i.e., travel with $v > c$ in the vacuum) in the electromagnetic case; 
and
are ``super-sonic" (i.e., travel with a speed larger than the sound-speed
in the medium) in the acoustic case. The first authors to produce
X-shaped waves {\em experimentally} were Lu \& Greenleaf[7] for
acoustics,
Saari et al.[7] for optics, and Mugnai et al. for microwaves[7].\\
 
\h In a recent paper of ours, appeared in this journal[8], we
showed that solutions to the Maxwell equations exist, that
displace themselves with Superluminal speed even along a normal
waveguide: where one ordinarily expects to meet propagating,
subluminal modes only.  Actually, a segment of ``undersized"
waveguide constitutes an evanescence region[9], and evanescent
waves are known to travel Superluminally[5,9-11]; however, it was
rather unexpected that (localized) waves could propagate
Superluminally down a normal-sized waveguide. In fact, the
dispersion relation in undersized guides is \ $\om^2/c^2 -
\beta^2 = - K^2$, \ so that the standard formula \ $v \simeq \drm
\om / \drm \beta$ \ yields a $v > c$ group-velocity[12]; by
contrast, in normal guides the dispersion relation becomes \
$\om^2/c^2 - \beta^2 = + K^2$, \ so that it seems to yield values
$v < c$ only. \ Instead, in our paper[8] we have shown that
localized solutions to Maxwell equations do exist, propagating
with $v > c$ even in normal waveguides; but their group-velocity
$v$ cannot be given$^{\# 1}$ by the approximate formula \ $v
\simeq \drm \om / \drm \beta$. \ [Let us recall that the
group-velocity is well defined only when the pulse has a clear
bump in space; but it can be calculated by the approximate,
elementary relation \ $v \simeq \drm \om / \drm \beta$ \ only
when $\om$
as a function of $\beta$ is also clearly bumped)].\\

{\bf 2. -- The infinite-energy solutions}\\
 
\h Namely, in ref.[8] we construced localized solutions to the Maxwell
equations (which propagate undistorted, with Superluminal speed along a
{\em cylindrical} waveguide located along the $z$-direction) for the
TM (transverse magnetic) case and for a dispersion-free medium. \
The case with dispersion has been treated
elsewhere[13], as well as the case of a co-axial cable[14]. \ Here, let
us
call attention to two points, which received just a mention
in ref.[8], with regard to eq.(9) and Fig.2 therein: \ (i) those
solutions
consist in {\em trains} of pulses (similar to the one depicted in Fig.2
of ref.[1]); \ (ii) each of such pulses is {\em $\X$-shaped}: See our
Fig.1 below. \ Let us
notice, incidentally, that we are referring ourselves to the
electromagnetic
case, but the same would hold for all situations in which a fundamental
role is played by the wave equation (as in acoustics, geophysics,
gravitational wave physics, etc.).
 
\h For intance, in the case of axial symmetry, let us consider a metallic
waveguide with radius[8]
 
 
$$r \equiv R\; .$$
 
Let us also put $\rhobf
\equiv (x,y)$, and $\rho = |\rhobf|$ and the bondary condition
$\Psi(\rho=r,z;t)=0$. In the previous paper[8] we constructed the
following solution,
 

\
 
\bb \Psi(\rho,z;t) \ugg \sum_{n=1}^{N} \left( \dis{{{2} \over
{a^2\,\sin^2\theta J_1^2(\la_n)}}} \right) \; J_0(K_n\,\rho) \;
\cos\left[\frac{\om_n}{V} \, (z - V\,t) \right] \ ,
\ee
 
\
 
where $\Psi$ represents the longitudinal component of the electric field,
$E_z$, while $N$ is an integer, the quantities $\la_n$ are the roots of
the Bessel function, $K_n=\la_n/R$, $\om_n=K_n\,c/\sin\theta$ and
$V=c/\cos\theta$. These solutions are therefore Fourier-Bessel--type
sums over different propagating modes with angular frequencies $\om_n$.
One can moreover notice that, in eq.(1), quantity
$\th$ is an arbitrary angle: by varying it, one obtains different
train speeds and different distance between the pulses. \ Actually, our
solutions propagate rigidly down the guide with (Superluminal) speed
$V = c / {\cos\th}$. \ In Fig.1 we depict one of the trains of X-shaped
waves, obtained by numerical evaluation of eq.(1) for a waveguide radius
$R = 5 \,$cm, with $\theta = \pi / 3$ (and, therefore, group-velocity
$V = 2 \, c$).
 
It is interesting to mention also
that the integer $N$ determines the space-time width of
the pulses: the higher $N$ is, the smaller the pulse ``spatio-temporal"
width
will be. \ Let us emphasize that each eq.(1) represents a
{\em multimodal} (but {\em localized}) propagation, as if the geometric
dispersion compensated for the multimodal dispersion.

\h We mentioned that $\Psi$ represents the electric field component
$E_z$.
Let us add that, by following the procedure adopted by us in ref.[8],
the other eletromagnetic field components in the considered TM case
result to be
 
\
 
\bb
\Ebf_\bot \ug i \, \dis{ {V \over {V^2-1 }} \; \sum_{n=1}^{\infty} {c
\over
\om_n} \,
\nabf_\bot \Psi } \ , 
\ee
 
\
 
where
 
$$\dis{ {V \over {V^2-1 }} \; \equiv \; {{\cos \th} \over {\sin^2 \th}} }
\ ;$$
 
and
 
\
 
\bb
\Hbf_\bot \ug \dis{  \varepsilon_0 V } \; {{\hat{\zbf}} \wedge
\Ebf_\bot} \ . 
\ee
 
\
 
\h Equation (1) allows for a physical interpretation, which
suggests a very simple way to get it. Each pulse train is a sum
of the first $N$ modes of our expansion (and for each $N$ we get
a different train, at our choice), whose frequencies have been
suitably chosen as corresponding to the {\em intersections} of
the modal curves (i.e., the various branches of the
dispersion-relation) with the single {\em straight} line $\om = V
\,\beta$ whose slope depends on $\th$ only: see Fig.2. \ In such
a case, all the modes correspond to the same (Superluminal) {\em
phase-velocity} $V_\ph$, it being independent of the mode
index~$n$; but, when the phase-velocity is independent of the
frequency, it becomes the {\em group-velocity}, which is the
velocity ``tout court" of the considered pulse. \ Let us repeat
once more that we thus got (non-evanescent) solutions to the
Maxwell equations, which are waves propagating undistorted along
normal waveguides with Superluminal speed, {\em even if} in
normal-sized waveguides the dispersion relation for each mode,
i.e. for each term of our Fourier-Bessel expansion, is the
ordinary ``subluminal" one, \ $\om^2/c^2 - \beta^2 = +K^2)$. \
Let us repeat that, in fact, that the {\em global} velocity $v$
(or group-velocity $v_\grm \equiv v$) of the pulses corresponding
to eq.(1) is not to be evaluated by the ordinary formula $v_\grm
\simeq \drm \om / d\beta$, valid for quasi-monochromatic signals).
This is at variance with the common situation in optical and
microwave communications, when the signal is usually superimposed
to a carrier wave whose frequency is generally much higher than
the signal bandwidth. In {\em that} case the standard formula for
$v_\grm$ yields the correct velocity to deal with (e.g., when
propagation delays are studied). Our case, on the contrary, is
much more reminiscent of a baseband modulated signal, as those
studied in ultrasonics: the very concept of a carrier becomes
meaningless here, as the elementary ``harmonic" components have
widely different frequencies.
 
\ The fact that our Superluminal
solutions travel rigidly, down the waveguide, is at variance also with
what
happens for truncated (Superluminal) solutions[15,7], which travel almost
rigidly only along their finite ``field depth" and then abruptly decay.
 
\h It may be finally underlined that the coefficients in eq.(1) can be
varied so to keep the pulse spectrum inside the desired frequency range.
This point will be discussed again soon.\\

{\bf 3. -- The finite-energy solutions}\\
 
\h In this note, we have called attention to the fact that
solutions (1) are infinite trains of pulses, with {\em infinity}
energy.  This is not a real problem (plane-waves too have
infinite energy), provided that we are able to truncate them in
space and time withount destroying their good properties. We
shall go on following the previous assumptions: what we are going
to do holds, however, for both the TM and the TE case. \ Let us
{\em anticipate} that, in order to get finite total-energy
solutions, we shall have to replace each characteristic frequency
$\om_n$ [cf. eq.(1), or Fig.2] by a {\em small} frequency band
$\Delta \om$ centered at $\om_n$, always choosing the same
$\Delta \om$ independently of $n$. In fact, since all the modes
entering the Fourier-type expansion (1) possess the same
phase-velocity $V_\ph \equiv V = c/\cos\th$, each small bandwidth
packet associated with $\om_n$ will possess {\em the same
group-velocity} $v_\grm = c^2/V_\ph$, so that we shall have as a
result a wave whose {\em envelope} travels with the {\em
subluminal} group-velocity $v_\grm$.  {\em However}, inside that
subluminal envelope, one or more {\em pulses} will be travelling
with the dual ({\em Superluminal\/}) speed $V=c^2/v_\grm$. Such
well-localized peaks will have nothing to do with the ordinary
(sinusoidal) carrier-wave, and will be regarded as constituting
{\em the relevant} wave. \ When integrating each term of
expansion (1) over its corresponding frequency-band, one may
choose, e.g., Gaussiam spectra.
 
\h Before going on, let us mention that previous work
related to FTESs can be found ---as far as we know--- only in
refs.[16,14,17].

\ More formally, let us consider
our ordinary solutions for a metallic waveguide, written in the form
 
\
 
\hfill{$
\psi_{n}(\rho,z;t) \ug A_n \; R_n(\rho) \; \cos \left[ \be(\om) \; z -
\om t \right] \ ,
$\hfill}
 
\
 
where coefficients $A_n$ and functions $R_n$ are given by the
coefficients and the (transverse) functions entering eq.(1); namely:
 
\
 
\hfill{$ A_n \ug \dis{{{2} \over {a^2\,\sin^2\theta
J_1^2(\la_n)}}} \;\; \ \ \ \ \ \  R_n(\rho) \ug J_0 (K_n \rho) \;
; \ \ \ \ \ \ \ \ \ \ K_n \ug \dis{{\lambda_n \over R}} \ .
$\hfill}
 
\
 
Then, let us adopt the spectral functions
 
\
 
\bb
W_n \equiv \exp [-q^2(\om-\om_n)^2] \ , \label{spec}
\ee  
 
\
 
where the weight-parameter $q$ is always the same, so that $\Delta \om$
too
is independent of $n$ [in fact, it is $\Delta \om = 1/q$]; and where
 
$$
\om_n \equiv \dis{ {{K_{n} c} \over {\sin\th}} } \ .
$$
 
quantity $\sin\th$ having a fixed but otherwise arbitrary value.
Notice that the last relation implies the wavenumbers $\beta_{n}$
of the longitudinal waves to be given, in terms of the
corresponding $\omega_n$, by \ $\beta_{n} = \omega_n\cos\theta/c$.
\ We can construct FTESs, \ ${{\cal F}}(\rho,z;t)$, \ of the
type\footnote{When integrating over $\om$ from $-\infi$ to
$+\infi$ also the non-physical (non-causal) components could
contribute, that travel backwards in space[17,14]. But their
actual contribution is totally negligible, since the
weight-functions $W_n$ are strongly localized in the vicinity of
the $\om_n$-values (which are all positive: see, e.g., Fig.2). In
any case, one could integrate from $0$ to $\infi$ at the price of
increasing a little the mathematical complexity: we are
preferring the present formalism for simplicity's sake.}
 
\
 
\bb
{\cal F}(\rho,z;t) \ug \sum_{n=1}^N \; \int_{-\infi}^\infi \, \drm\om \,
\psi_{n} \, W_n \ , \label{F}   
\ee
 
\
 
with arbitrary $N$. Notice that we are not using a single Gaussian
weight,
but a different Gaussian function $W_n$ for each $\om_n$-value. Such
weights
$W_n$ are well localized around the corresponding $\om_n$, so that one
can
expand (for each value of $n$, in the above sum) the function $\be(\om)$
in the neighbourhood of the corresponding $\om_n$-value as follows:
 
$$
\be(\om) \; \simeq \; \be(\om_n) \, + \, \dis{ \left. {{\pa\be}
\over {\pa\om}} \right|_{\om_n} } (\om - \om_n) + \ ...
$$
 
where $\be(\om_n) = \om_n \cos\th/c$, and the further terms are
neglected [since, let us repeat, $\Delta \om$ has been assumed to
be small]. Therefore, we are now facing no longer a set of
phase-velocities, but the set of group-velocities
 
$$ \dis{ {1 \over {v_{\grm n}}} \ugg  \left. {{\pa\be} \over {\pa\om}}
\right|_{\om_n} } \ , $$
 
which result to be {\em independent} of $n$, all of them possessing
therefore the same value
 
\
 
\hfill{$ \dis{ v_{\grm n} \, \equiv \, v_\grm \ug c\,{\cos\th} } \
. $\hfill} (6)
 
\
 
By performing the integration in eq.(5), we eventually obtain
 
\
 
\hfill{$ {\cal F}(\rho,z;t) \ugg \dis{ {\pisr \over q} \exp
\left[ -{{(z-v_\grm\,t)^2} \over {4q^2\,v^2_\grm}} \right] }
\Psi(\rho,z-Vt) \ , $\hfill} (7)
 
\

where $\Psi(\rho,z-Vt)$ is the pulse train given by eq.(1); and we had
recourse to the identity
 
$$
\int_{-\infi}^{\infty} \drm f \; \exp[-q^2 f^2] \; \cos[f
(z-v_\grm\,t)/v_\grm] \ugg \dis{ {\pisr \over q} \exp \left[
-{{(z-v_\grm\,t)^2} \over {4q^2v^2_\grm}} \right] } \ .
$$
 
\h It is rather interesting that our FTESs are related to the
X-shaped waves, since eq.(5) has been written in the form (7), where
the function $\Psi(\rho,z-Vt)$ is any one of our previous solutions in
eq.(1)
above, at our choice.\\

{\bf 4. -- Conclusions}\\
 
\h In conclusion, looking for {\em finite} total-energy
solutions, we have found a Gaussian envelope that travels with a
{\em subluminal} velocity $v=c\,\cos\th$. However, inside it, we
have got a train of pulses travelling Superluminally (with $V =
c^2 /v = c/\cos\th$). And we can control the number of pulses
inside the envelope just by varying the value of $q$.
 
\h We have actually shown that, if we choose the $\om_n$-values
as in Fig.2, all the small-bandwith packets centered in the
$\om_n$'s will have the same phase-velocity $V>c$ and therefore
the same group-velocity $v_\grm < c$ [since for metallic
waveguides the quantities $K_n^2=\om_n^2/c^2-\be^2$ are constant
for each mode, and $v_\grm \equiv \pa\om / \pa\be$, so that it is
$V \, v_\grm = c^2$].  This means that the envelope of solution
(5)-(7) moves with slower-than-light speed; the envelope
length\footnote{One may call ``envelope length" the distance
between the two points in which the envelope heigth is, for
instance, 10\% of its maximum heigth.} $\Delta \ell$ depending on
the chosen $\Delta \om$, and being therefore proportional to $q
v_\grm$. However, inside such an envelope, one has a train of
(X-shaped) pulses
---having nothing to do with the ordinary carrier wave,
\footnote{Actually, they can be regarded as a sum of carrier waves.}---
travelling with the Superluminal speed $V$.
 
\h An interesting
point is that we can choose the envelope length so that it contains
{\em only one} (X-shaped wave) peak. Even if the (global) speed of the
envelope is subluminal in the finite-energy case, while Superluminal
speeds
are met only {\em locally} (internally), nevertheless in the present case
the Superluminal speed $V=c^2/v_\grm$ of such a ``single" pulse might be
regarded as the actual velocity of the wave. \ In order to have just one
peak inside the envelope, the
envelope length is to be chosen smaller than the distance between two
successive peaks of the (infinite total energy) train (1). \ The
amplitude of such a single X-shaped
pulse (which remains confined inside the envelope boundary) first
increases,
and afterwards decreases, while travelling; till when it practically
disappears. While the considered pulse tends to vanish on the right
(i.e., under the right tail of the envelope), a second pulse starts to be
created
on the left; and so on [from eq.(7) it is clear, in fact, that our
finite-energy solutions are nothing but an (infinite-energy) solution of
the type in eq.(1), multiplied by a Gaussian function]. \ We illustrate
such a behaviour in Figs.3, namely, in the set of eight figures from Fig.3a
to Fig.3h. \ We have depicted a similar
behaviour in the last set of figures of ref.[13], when studying the case
of a co-axial guide.

\h Let us finally remark that similar considerations could be extended
to all the situations where a waveguide supports several modes. Tests at
microwave frequencies, for instance, should be rather easy to perform; by
contrast, experiments in the optical domain are made difficult, at
present,
by the limited extension of the spectral windows corresponding to not too
large attenuations: we shall discuss this point elsewhere.
\h It is rather interesting that our FTESs are related to the
X-shaped waves, since in eq.(7) the function $\Psi(\rho,z-Vt)$ is
any one of our previous solutions in eq.(1) above, at our choice.\\
 
\
 
\
 
{\bf Acknowledgements}\\
 
\h The authors are very grateful to Hugo E.Hern\'andez-Figueroa and
K.Z.N\'obrega (Fac. of Electric Engineering, Unicamp), and to Amr Shaarawi
(Cairo University). Useful discussions are moreover acknowledged
with T.F.Arecchi and C.Dartora, as well as with V.Abate, A.Attiya, F.Bassani,
C.Becchi, M.Brambilla, C.Cocca, R.Collina, R.Colombi, C.Conti, G.C.Costa,
P.Cotta-Ramusino, G.Degli Antoni,
R.Garavaglia, L.C.Kretly, G.Kurizki, D.Mugnai, G.Privitera, V.Petrillo,
A.Ranfagni, A.Salanti, G.Salesi, J.W.Swart, M.T.Vasconselos
and M.Villa.

\vfill\newpage
 
\centerline{{\bf Figure Captions}}
 
\
 
{\bf Fig.1 --} This figure depicts one of our (infinite total-energy)
localized solutions, given in eq.(1). It consists, as expected, in a
{\em train} of X-shaped waves; and propagates rigidly along the
considered, normal-sized circular waveguide, with radius $r = 5 \,$cm,
with the Superluminal speed $v = c / \cos\theta$.  In this figure the
chosen value of $\theta$ is $\theta = \pi/3$.\\
 
{\bf Fig.2 --} Dispersion curves for the symmetrical $\TM_{0l}$ modes in
a perfect cyclindrical waveguide, and location of the frequencies whose
corresponding modes have equal phase-velocity. See the text.\\
 
{\bf Figs.3 --} This set of figures depicts one of our {\em finite}
total-energy localized solutions, given in eqs.(5) and (7).
Indeed, they show the time evolution of a finite total energy solution:
Choosing $q=2.041 \times 10^{-10}$ and $\theta=\pi / 3$ (and normalized
units), there is only {\em one} X-shaped pulse inside the subluminal
envelope: see the text. The pulse and envelope velocities are given by
$V=1/cos\theta$ and $v_{g}=1/V$. The (global) speed $v$ of the envelope
is therefore subluminal in the finite-energy case, while Superluminal speeds
$V$ are met only {\em locally} (internally). Nevertheless in the present case
the Superluminal speed $V=c^2/v_\grm$ of such a ``single" pulse 
could be regarded as the actual velocity of the wave. \ Figures (a), (b),
(c), (d), (e), (f), (g) and (h) show a complete cycle of the pulse;
they correspond to the time instants written at the top of each of them.

\vfill\newpage
 
\centerline{{\bf References}}
 
\
 
[1] H.Bateman: {\em Electrical and Optical Wave
Motion}  (Cambridge Univ.Press; Cambridge, 1915).\hfill\break
 
[2] A.O.Barut and A.Grant:  {\em Found. Phys. Lett.}
3 (1990) 303;  A.O.Barut and A.J.Bracken: {\em Found. Phys.} 22 (1992)
1267. \ See also refs.[14,19,21] below.\hfill\break
 
[3] A.O.Barut and H.C.Chandola:  {\em Phys. Lett.}
A180 (1993) 5. \ See also A.O.Barut: {\em Phys. Lett.} A189 (1994)
277, and A.O.Barut et al.: refs.[3].\hfill\break
 
[4] R.Donnelly and R.W.Ziolkowski: {\em Proc. Royal Soc. London} A440
(1993) 541; \ I.M.Besieris, A.M.Shaarawi and R.W.Ziolkowski:
{\em J. Math. Phys.} 30 (1989) 1254]; \ S.Esposito: {\em Phys.
Lett.} A225 (1997) 203; \ W.A.Rodrigues Jr. and J.Vaz Jr., {\em Adv.
Appl.
Cliff. Alg.} S-7 (1997) 457.\hfill\break
 
[5] See, e.g., E.Recami: ``Classical tachyons and possible applications,"
{\em Rivista Nuovo Cimento} 9(6) (1986), issue no.6 (pp.1-178); and
refs. therein.\hfill\break
 
[6] Jian-yu Lu and J.F.Greenleaf: {\em IEEE Transactions on Ultrasonics,
Ferroelectrics, and Frequency Control} 39 (1992) 19; \  R.W.Ziolkowski,
I.M.Besieris and A.M.Shaarawi: {\em J. Opt. Soc. Am.} A10 (1993) 75; \
E.Recami: {\em Physica} A252 (1998) 586.\hfill\break
 
[7] Jian-yu Lu and J.F.Greenleaf: {\em IEEE Transactions on Ultrasonics,
Ferroelectrics, and Frequency Control} 39 (1992) 441; \  P.Saari and
K.Reivelt: {\em Phys. Rev. Lett.} 79 (1997) 4135; \ D.Mugnai, A.Ranfagni
and R.Ruggeri: {\em Phys. Rev. Lett.} 84 (2000) 4830.\hfill\break
 
[8] M.Zamboni Rached, E.Recami and F.Fontana: {\em Phys. Rev.} E64 (2001)
066603, available also as e-print physics/0001039. \ Cf. also Jian-yu Lu:
(1992, unpublished).\hfill\break
 
[9] Cf., e.g., A.M.Steinberg, P.G.Kwiat and R.Y.Chiao: {\em Scient. Am.}
269 (1993) issue no.2, p.38; A.Enders and G.Nimtz: {\em J. de Physique-I}
3 (1993) 1089; {\em Phys. Rev.}  E48 (1993) 632.\hfill\break
 
[10] Cf., e.g., A.P.L.Barbero, H.E.Hern\'andez-Figueroa and E.Recami:
{\em Phys. Rev.} E62 (2000) 8628, and refs. therein; \ H.M.Brodowsky,
W.Heitmann and G.Nimtz: {\em Phys. Lett.} A222 (1996) 125.\hfill\break
 
[11] V.S.Olkhovsky and E.Recami: {\em Phys. Reports} 214 (1992) 339,
and refs. therein; \ V.S.Olkhovsky, E.Recami, F.Raciti and A.K.Zaichenko:
{\em J. de Physique-I} 5 (1995) 1351; \ T.E.Hartman: {\em J. Appl. Phys.}
33 (1962) 3427.\hfill\break
 
[12] E.Recami, F.Fontana and G.Garavaglia: {\em Int. J. Mod. Phys.} A15
(2000) 2793; \ R.Garavaglia: Thesis work (Dip. Sc. Informazione,
Universit\`a statale di Milano; Milan, 1998; G.Degli Antoni and
E.Recami supervisors).\hfill\break
 
[13] M.Z.Rached, K.Z.Nobrega, E.Recami \& H.E.Hern\'andez F.: ``Localized
Superluminal solutions to the wave equation in (vacuum or) dispersive
media, for arbitrary frequencies and with adjustable bandwidth",
submitted for pub. \hfill\break
 
[14] M.Z.Rached, K.Z.Nobrega, E.Recami \& H.E.Hern\'andez F.:
``Superluminal X-shaped beams propagating without distortion along a
co-axial guide", to appear in {\em Phys. Rev. E}. \hfill\break
 
[15] R.W.Ziolkowski, I.M.Besieris and A.M.Shaarawi: {\em J. Opt. Soc.
Am.} A10 (1993) 75; \ E. Recami: ref.[6].\hfill\break
 
[16] I.M.Besieris, M.Abdel-Rahman, A.Shaarawi and A.Chatzipetros:
Progress in Electromagnetic Research (PIER) 19 (1998) 1-48
(1998).\hfill\break
 
[17] M.Zamboni Rached, E.Recami and H.E.Hern\'aqndez-Figueroa: ``New
localized Superluminal solutions to the wave equations with finite
total energies and arbitrary frequencies", in press in {Europ. Phys.
Journal - D}.\hfill\break

\end{document}